\documentclass{article}
\usepackage{amsmath,graphicx,mlspconf}

\usepackage{amsfonts}
\usepackage{amssymb, color, bm}

\usepackage{algorithmicx}
\usepackage{algpseudocode}
\usepackage{algorithm,setspace}
\usepackage{xcolor}
\usepackage{physics}

\usepackage{appendix}

\DeclareMathOperator{\E}{\mathbb{E}}

\title{Quantum-Aided Meta-Learning for Bayesian Binary Neural Networks via Born Machines}
\name{Ivana Nikoloska and Osvaldo Simeone \thanks{E-mails: \{ivana.nikoloska, osvaldo.simeone\}@kcl.ac.uk.. This work was supported by the European Research Council (ERC) under the European Union’s Horizon 2020 Research and Innovation Program (Grant Agreement No. 725731).}}
\address{KCLIP, CTR, Department
of Engineering, King's College London}

\begin{document}

\maketitle

\begin{abstract}
Near-term noisy intermediate-scale quantum circuits can efficiently implement  implicit probabilistic models in discrete spaces, supporting distributions that are practically infeasible to sample from  using classical means.  One of the possible applications of such models, also known as Born machines, is probabilistic inference, which is at the core of Bayesian methods. This paper studies the use of Born machines for the problem of training binary Bayesian neural networks. In the proposed approach, a Born machine is used to model the variational distribution of the binary weights of the neural network, and data from multiple tasks is used to reduce training data requirements on new tasks. The method combines gradient-based meta-learning and variational inference via Born machines, and is shown in a  prototypical regression problem to outperform conventional joint learning strategies.
\end{abstract}%

\section{Introduction}
\label{sec:intro}

\begin{figure}[tbp]
\centering
\includegraphics[width=0.4\textwidth]{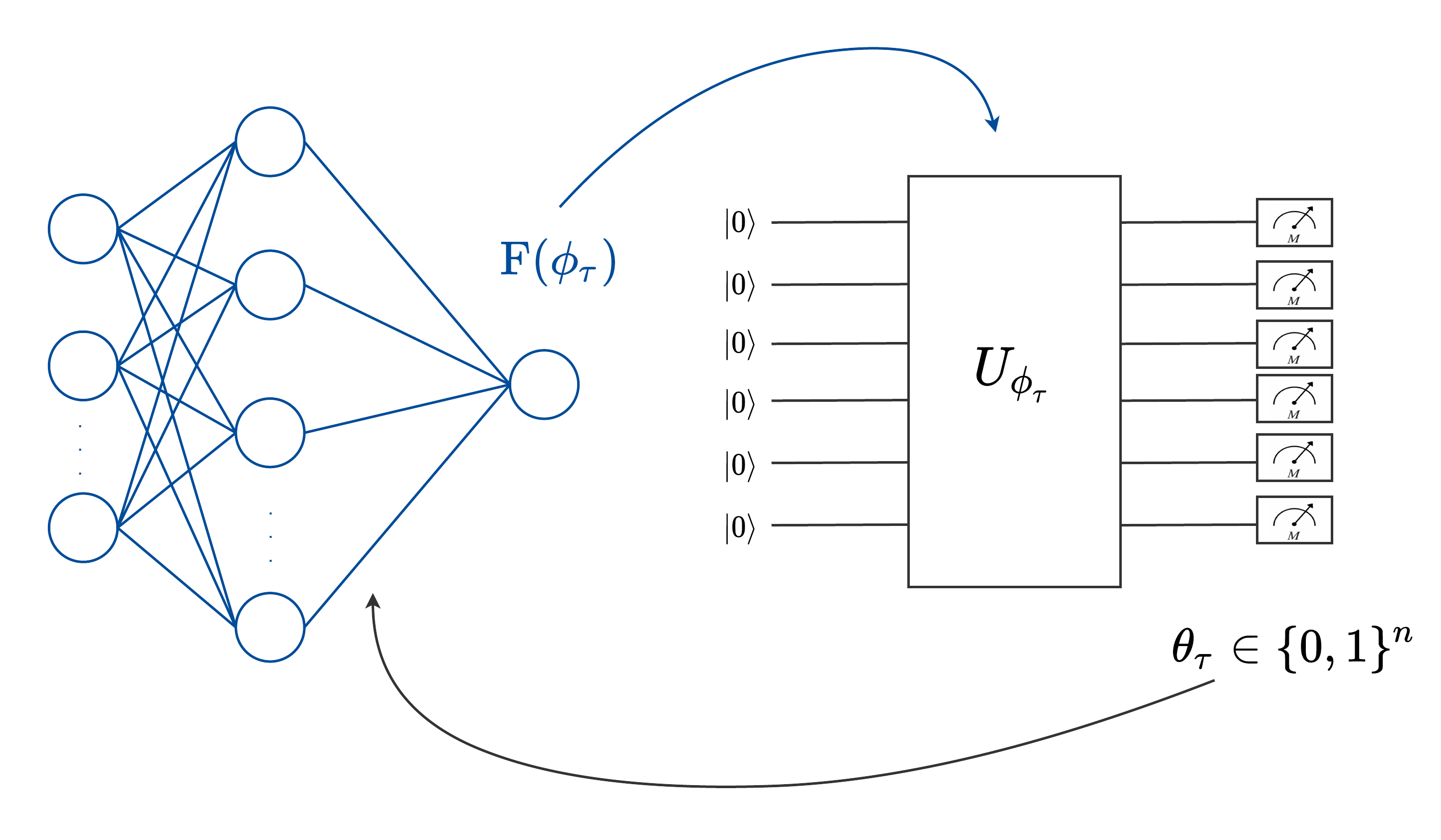}\vspace{-0.4cm}
\caption{(left) A  binary Bayesian neural network, i.e., a neural network with stochastic binary weights, is trained to carry out a learning task. (right) The probability distribution of the binary weights of the neural network is modelled by a Born machine, i.e., by a parametric quantum circuit (PQC), leveraging the PQC's capacity to model complex distributions \cite{arute2019quantum}. The goal of this paper is to develop meta-learning strategies that are able to quickly adapt the parameters of the Born machine so as to enable fast adaption to new learning tasks. }
\vspace{-0.5cm}
\label{sys_one}
\end{figure}

Learning a new task from a few examples inherently induces a significant amount of uncertainty. A principled framework to reason about uncertainty is given by Bayesian learning methods \cite{gal2016dropout, louizos2017multiplicative}, which encode epistemic uncertainty into a distribution over the models of a given model class. A typical example, illustrated in Fig. \ref{sys_one}, is the training of Bayesian neural networks, which results in a distribution over the neural weights.  When the model parameters are discrete, representing and optimizing distributions is particularly challenging  \cite{andriyash2018improved}. This paper explores the use of quantum generative models -- also known as \textbf{Born machines} \cite{coyle2020born, wang2018boltzmann} -- as a potentially more efficient alternative to standard classical models for the problem of training \textbf{binary Bayesian neural networks}. A novel method is proposed that integrates \textbf{meta-learning} with the gradient-based optimization of quantum Born machines \cite{liu2018differentiable}, with the aim of speeding up adaptation to new learning tasks from few examples.



Meta-learning transfers information from previously observed tasks to as of yet unknown, new, tasks in the form of an inductive bias, with the goal of improving sample efficiency on novel tasks \cite{pratt1998learning}. In a classical frequentist setting, the  inductive bias can be optimized in several different ways, such as by tuning hyperparameters that regularize task-specific parameters \cite{heskes1998solving} or by designing an optimization algorithm \cite{finn2017model}. \textbf{Bayesian meta-learning} can improve the sample efficiency of Bayesian learning by optimizing the inductive bias through the prior distribution -- an approach also known as empirical Bayes -- or through the hyperparameters of the algorithm used to minimize the variational free energy  \cite{amit2018meta}.

The integration of meta-learning and quantum machine learning has been recently explored in \cite{wilson2021optimizing, verdon2019learning, huang2022learning}. These works rely on frequentist meta-learning to speed up training of parametric quantum models, such as the variational quantum eigensolver (VQE) via classical machine learning models. In this regard, this paper considers the opposite setting, in which quantum circuits -- Born machines -- are used to facilitate the training of classical binary neural networks. In order to enable optimization in the discrete space of binary weights, we adopt a Bayesian meta-learning formulation aided by Born machines. The proposed approach applies the variational inference method introduced in \cite{benedetti2021variational} to the problem of Bayesian meta-learning.

The rest of this paper is organized as follows. Sec. 2 introduces the setting and problem definition. Sec. 3 presents the proposed Bayesian meta-learning solution, which is evaluated in Sec. 4. The paper is concluded in Sec. 5.





\section{Setting and Problem Definition}
\label{sec:ml_b}

\subsection{Setting}
As illustrated in Fig. \ref{sys_one}, we are interested in training \textbf{Bayesian binary neural networks}, i.e., classical neural networks with stochastic binary weights, in a sample-efficient manner by means of meta-learning. Binary neural networks facilitate hardware implementation \cite{qin2020binary}, and Bayesian neural networks can faithfully represent predictive uncertainty via the incorporation of prior knowledge and ensembling \cite{mullachery2018bayesian}. The key idea of this work is to model the distribution $q(\theta)$ of the binary weights $\theta$ via a Born machine, i.e., via a probabilistic parametric quantum circuit (PQC), due to the capacity of PQCs to efficiently implement complex probability distributions   \cite{arute2019quantum,sweke2021quantum}. We adopt a classical meta-learning formulation, in which one has access to data from multiple learning tasks, and the goal is to optimize a procedure that can optimize distribution  $q(\theta)$ from few examples related to a new learning task.

To formalize the setting, we assume (offline) access to \textbf{meta-training data} for a set of $\mathcal{T}$ \textbf{meta-training tasks}. For each meta-training task $\tau\in\{1,...,\mathcal{T}\}$, we have a training set $\mathcal{D}_{\tau}^{\text{tr}}$ and a test set $\mathcal{D}_{\tau}^{\text{te}}$, with an overall number of examples $N_{\tau}$. Training is applied \emph{separately} to each task $\tau$, producing a distribution $q(\theta_\tau)$ on the binary parameters $\theta_{\tau}$ as a function of the training data $\mathcal{D}_{\tau}^{\text{tr}}$ and of a  hyperparameter vector $\xi$. The real-valued hyperparameter vector $\xi$ is \emph{shared} across tasks, and it determines the training process that is applied \emph{separately} to each task. The goal of meta-learning is to optimize the hyperparameters $\xi$ based on the available meta-training data, so as to ensure that the distribution $q(\theta_{\tau^{*}})$ optimized based on few data points for a new task $\tau^{*}$ provides satisfactory test loss.
 
We follow \textbf{variational inference (VI)} by parametrizing the distribution $q(\theta_\tau)$ via a real-valued vector of parameters $\phi_\tau$ for each task $\tau$. We write the resulting parametrized variational distribution as $q_{\phi_\tau}(\theta_\tau)$. As detailed in the next section, the distribution $q_{\phi_\tau}(\theta_\tau)$ is implemented by a PQC, with vector $\phi_\tau$ representing the parameters of the PQC. Training for each task $\tau$ amounts to the optimization of the  \textbf{variational parameters} $\phi_\tau$, and hence of the distribution $q_{\phi_\tau}(\theta_\tau)$, as a function of the training data $\mathcal{D}_{\tau}^{\text{tr}}$ and of a  hyperparameter vector $\xi$. 

To define the training problem for each task $\tau$, we follow (generalized) \textbf{Bayesian learning}. Let us denote as $p(x| \, \theta_\tau)$ the probability distribution assigned by the binary neural network to data point $x$ given model parameters $\theta_\tau$. Note that one can easily consider discriminative models by introducing a conditioning over covariates (see Sec. 4). We also write as $p(\mathcal{D}_\tau^\text{tr} \, | \, \theta_\tau)=\prod_{x\in \mathcal{D}_\tau^\text{tr}}p(x| \, \theta_\tau) $ the corresponding probability assigned to the training set $\mathcal{D}_\tau^\text{tr}$ for task $\tau$. Given a prior distribution $p(\theta_\tau)$, the training criterion for each task $\tau$ is given by the the \textbf{free energy} 
\begin{align}\label{ELBO}
    \textrm{F}(\phi_\tau) = -\E_{q_{\phi_\tau}(\theta_\tau)}   [\log p(\mathcal{D}_\tau^\text{tr} \, | \, \theta_\tau)]  
     + \beta \, \textrm{KL}(q_{\phi_\tau}(\theta_\tau) \, || \, p(\theta_\tau)),
\end{align}
with $\textrm{KL}(\cdot \, || \, \cdot)$ denoting the Kullback-Leibler (KL) divergence; and $\beta\geq 0$ being a regularization coefficient. The free energy defines a trade-off between data fitting, which is expressed by the average training log-loss in the first term, and deviation of the distribution $q_{\phi_\tau}(\theta_\tau)$ from the prior, which is measured by the second, regularization, term. The variational parameter vector $\phi_\tau$ is obtained during training by addressing the minimization of the free energy
\begin{align}\label{phi_min}
    \phi_\tau(\xi) \underset{\xi}{\leftarrow} \underset{\phi_\tau}{\text{ arg min}} \,\,\, \textrm{F}(\phi_\tau).
\end{align}
 In \eqref{phi_min}, the notation indicates that the optimal $\phi_\tau(\xi)$ is generally a function of the hyperparameters  $\xi$. For example, as we will explore in Sec. 3, the solution may be obtained via gradient descent with initialization $\xi$.

At the level of meta-learning, the hyperparameter $\xi$ is obtained by minimizing the average test log-loss on the the meta-training tasks (see, e.g., \cite{yoon2018bayesian}), i.e., 
\begin{align}\label{eq:bilevel}
    \xi \leftarrow \underset{\xi}{\text{ arg min}} \,\,\, -\frac{1}{\mathcal{T}} \sum_{\tau = 1}^\mathcal{T} \log p(\mathcal{D}_\tau^\text{te} \, | \, \theta_\tau(\xi)),
\end{align}where the model parameters $\theta_\tau(\xi)$ depend on the hyperparameters $\xi$ via (\ref{phi_min}). As we will see in Sec. 3, this optimization can also be implemented via gradient descent.

\begin{figure}[tbp]
\centering
\includegraphics[width=0.99\linewidth]{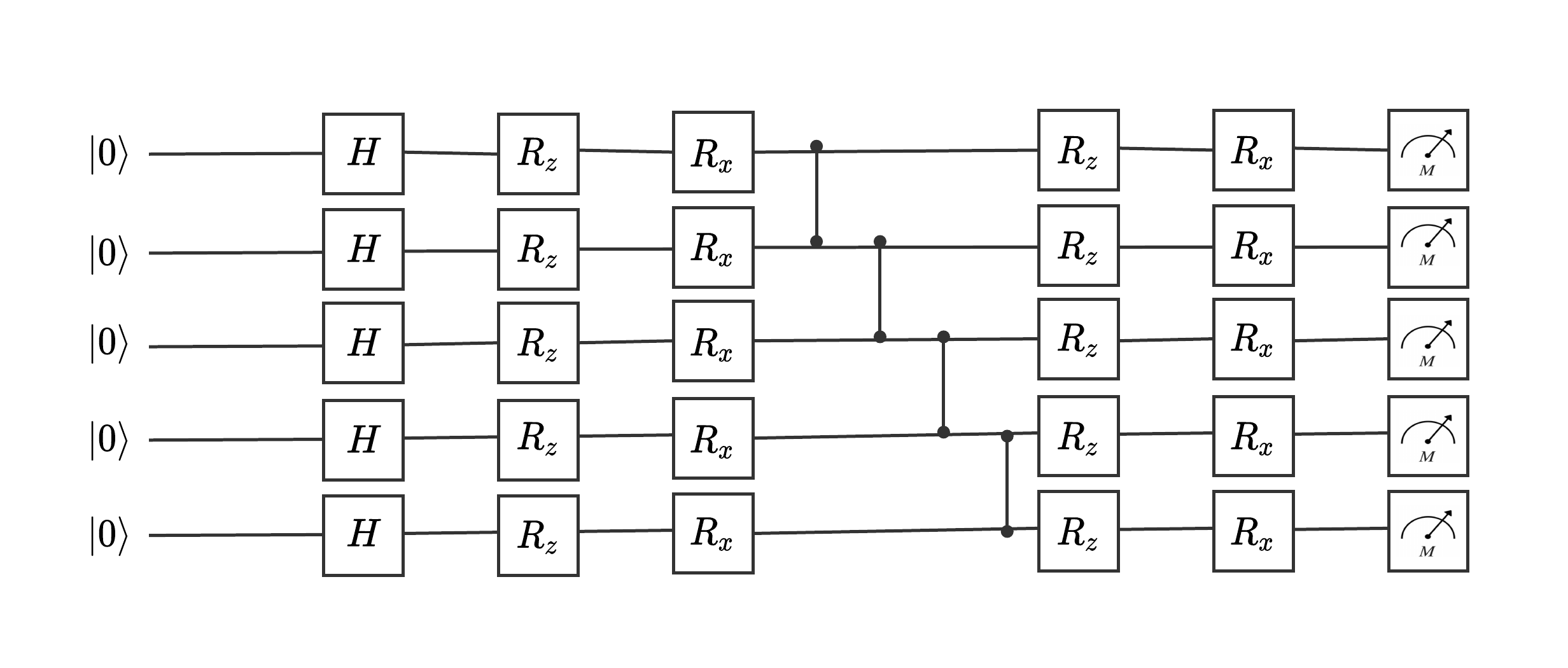}
\caption{Hardware-efficient ansatz for the Born machine used in the experiments. All qubits are initialized in the ground state $\ket{0}$. The rotations $R_x$, $R_z$ are parametrized by the entries of the variational vector $\phi_\tau$.}
\vspace*{-5mm}
\label{circ_one}
\end{figure}

\subsection{Born Machines as Implicit Variational Distributions}
In this paper, we model the distribution $q_{\phi_\tau}(\theta_\tau)$ by using a Born machine. The \textbf{Born machine} produces random binary strings $\theta_\tau \in\{0,1\}^{n}$, where $n = |\theta_\tau|$ denotes the total number of model parameters,  by measuring the output of a PQC $U_{\phi_\tau}$  defined by parameters $\phi_\tau$. As illustrated in Fig. \ref{circ_one}, the PQC takes the initial state $\ket{\psi_0} = \ket{0}^{\otimes n}$ of $n$ qubits as an input, and operates on it via a sequence of unitary gates described by a unitary matrix $U_{\phi_\tau}$. This operation outputs the final quantum state \begin{equation}\ket{\psi_{\phi_\tau}} = U_{\phi_\tau} \ket{\psi_0},\end{equation} which is measured in the computational basis to produce a random binary string $\theta_\tau \in\{0,1\}^{n}$. Note that each basis vector of the computational basis corresponds to one of all the possible $2^n$ patterns of model parameters $\theta_\tau$.

The PQC is implemented here using a \textbf{hardware-efficient ansatz} \cite{kandala2017hardware}, in which a layer of one-qubit unitary gates, parametrized by vector $\phi_\tau$, is followed by a layer of fixed, entangling, two-qubit gates. This pattern can be repeated any number of times, building a progressively deeper circuit. 

By Born's rule, the probability distribution of the output model parameter vector $\theta_\tau$ is given by
\begin{align}\label{pqs}
    q_{\phi_\tau}(\theta_\tau) = |\bra{\theta_\tau}\ket{\psi_{\phi_\tau}}|^2.
\end{align} Importantly, Born machines only provide samples via (\ref{pqs}), while the actual distribution (\ref{pqs}) can only be estimated by averaging multiple measurements of the PQC's outputs. Therefore, Born machines model \textbf{implicit distributions}, and only define a stochastic procedure that directly generates samples. 



\section{Bayesian Meta-Learning via Born Machines}
In this section, we introduce a gradient-based method to address the bi-level optimization problem defined by  \eqref{eq:bilevel}.  To this end, we assume that the inner optimization (\ref{phi_min}) is carried out for each meta-training task $\tau$ via gradient descent starting from a shared initialization given by the hyperparameter vector $\xi$.  Similar approaches have been used in \cite{amit2018meta, hu2020empirical}.  However, such works consider  continuous model parameters $\theta_\tau$, and use \emph{explicit}, parametric, variational distribution $q_{\phi_\tau}(\theta_\tau)$, such as Gaussian distributions. In contrast, this work assumes binary model parameters $\theta_\tau$, as well as an implicit variational distribution  \eqref{pqs} through the Born machine described in the previous section.


Using a single step of gradient descent for the inner training problem (\ref{phi_min}) on task $\tau$, we obtain the update
\begin{align}\label{update_phi}
    \phi_\tau (\xi) = \xi - \eta \nabla_{\xi} \textrm{F} (\xi),
\end{align}
where $\eta$ denotes the learning rate and we recall that the hyperparameter $\xi$ serves as initialization. The update in \eqref{update_phi} can be directly generalized to an arbitrary number $m\geq 1$ of gradient descent steps (see, e.g., \cite{finn2017model}), and we consider here a single step to simplify the presentation. Adddressing problem \eqref{eq:bilevel} via gradient descent yields the update
\begin{align}\label{update_theta}
    \xi &\leftarrow  \xi - \eta  \frac{1}{\mathcal{T}}\sum_{\tau=1}^{\mathcal{T}} \nabla_{\xi}(-\log p(\mathcal{D}_\tau^\text{te} \, | \, \theta_\tau(\xi))),
\end{align}
where $\eta$ denotes the learning rate. As summarized in Algorithm 1, at each meta-training iteration, the outer update (\ref{update_theta}) is carried out by summing the gradients over a mini-batch of tasks. Furthermore, the inner updates (\ref{update_phi}) are applied to the selected mini-batch of tasks within an inner loop consisting of $m$ updates.


\subsection{Computing the Gradients via Density-Ratio Approximation}
The inner-loop update (\ref{update_phi}) require the computation of the gradient $\nabla_{\xi} \textrm{F} (\xi)$ of the free energy. 
Rewriting the free-energy in \eqref{ELBO} using the definition of the KL divergence, i.e., 
\begin{align}\label{ELBO_re}
    \textrm{F}(\phi_\tau) =&  -\E_{q_{\phi_\tau}(\theta_\tau)}   [\log p(\mathcal{D}_\tau^\text{tr} \, | \, \theta_\tau)]  \nonumber\\
    & + \beta \,\, \E_{q_{\phi_\tau}(\theta_\tau)} \left[ \log \left(\frac{q_{\phi_\tau}(\theta_\tau)}{p(\theta_\tau)}\right) \right],
\end{align}
reveals the need to compute the density ratio between the variational distribution $q_{\phi_\tau}(\theta_\tau)$ and the prior $p(\theta_\tau)$. However, as mentioned, the considered Born machine-based model for distribution $q_{\phi_\tau}(\theta_\tau)$ is implicit, making it difficult to compute this term.
To address this challenge, we adopt an approximation that uses the \textbf{density-ratio estimation} method proposed in \cite{mohamed2016learning}, and also used in \cite{benedetti2021variational}.

Accordingly, the ratio in (\ref{ELBO_re}) is estimated by using the output of a binary classifier that is trained to recognize samples from the variational distribution $q_{\phi_\tau}(\theta_\tau)$ against samples from the prior $p(\theta_\tau)$. Specifically, we introduce a parametric classifier defined by a real-valued parameter vector $w$ that takes as input a vector $\theta_\tau\in\{0,1\}^n$ and outputs a probability $d_{w}(\theta_\tau)$. This probability provides an estimate of the probability that the sample was produced from the distribution $q_{\phi_\tau}(\theta_\tau)$. If the classifier is well trained, it is possible to show that one has the approximate equality \cite{mohamed2016learning}
\begin{align}\label{log-odds}
    \log \left(\frac{q_{\phi_\tau}(\theta_\tau)}{p(\theta_\tau)}\right) \approx \log\left( \frac{d_{w}(\theta_\tau)}{1-d_{w}(\theta_\tau)}\right) = \text{logit} (d_{w}(\theta_\tau)),
\end{align}where we have introduce the logit, or odds, produced by the classifier. 

The approximate equality (\ref{log-odds}) provides the density-ratio estimate. Accordingly, using $K$ samples $\{\theta_\tau^k\}_{k=1}^{K}$ from distribution $q_{\phi_\tau}(\theta_\tau)$ generated by the Born machine to approximate the expectations in \eqref{ELBO_re}, the estimate of the free energy is obtained as 
\begin{align}\label{ELBO_re_fin}
    \textrm{F}(\phi_\tau)  \approx & -\frac{1}{K} \sum_{k=1}^K  \log p(\mathcal{D}_\tau^\text{tr} \, | \, \theta_\tau^k)  
    + \beta \,\, \frac{1}{K} \sum_{k=1}^K \left( \text{logit} (d_{w}(\theta_\tau^k)) \right).
\end{align}

The classifier can be in practice trained using stochastic gradient descent by tackling the minimization of the cross-entropy \begin{align}\label{lossG}
\textrm{G}(w) = &-\E_{p(\theta_\tau)}   [\log d_{w}(\theta_\tau)] 
-\E_{q_{\phi_\tau}(\theta_\tau)}  [ \log (1-d_{w}(\theta_\tau))]
\end{align} via updates of the form
 \begin{align}\label{update_w}
     w \leftarrow w + \eta \nabla_{w} \textrm{G}(w).
 \end{align}The optimization of the classifier and the ELBO is carried away in tandem as detailed in Algorithm 1. Further details on the computation of the gradients via the parameter shift rule \cite{schuld2018supervised} can be found in Appendix A. 

\begin{algorithm}
\caption{Meta-learning X Born machines}

\begin{algorithmic}[1]

\Procedure{Meta-training}{}       
    \State Initialise parameters $\theta$
    \For{$k \,\,\, \text{meta-training iterations}$}
        \State Select meta-training tasks $\tau \in \{1,...,\mathcal{T}\}$
        \For{$\text{all selected meta-training tasks} \,\,\, \tau$}
        \State $\phi_\tau = \xi$
        \For{$m \,\,\, \text{iterations}$}
        \State Draw $K$ samples $\theta_\tau \sim q_{\phi_\tau}(\theta_\tau \, | \, \mathcal{D}_\tau^{\text{tr}})$
        \State Optimize the classifier as in \eqref{update_w}
        \State Optimize the Born machine as in as in \eqref{update_phi}
        \EndFor
        
    \EndFor
\State Update initialization as in \eqref{update_theta}
\EndFor
\EndProcedure

\end{algorithmic}
\end{algorithm}

\section{Experiments}
\label{sec:exp}

In this section, we provide experimental results to elaborate on the performance of the proposed meta-learning scheme.

\noindent \textbf{Experimental Setup.} Reflecting the limited number of available qubits in state-of-the-art quantum hardware and emulators, we resort to a synthetic, prototypical, environment for regression.
Following \cite{finn2017model}, for each task $\tau$, the population distribution underlying the generation of training and test data produces input-output pairs $(x^n_\tau, y^n_\tau)$ such that 
\begin{align}
    x^n_\tau \sim \mathcal{U}(2, 7) \, \text{and} \, y^n_\tau \sim \mathcal{N}(\textrm{g}_{a,b,c,\alpha} (x^n_\tau),0.12),
\end{align}
where the function 
\begin{align}\label{sinus}
    \textrm{g}_{a,b,c,\alpha} (x^n_\tau) = \alpha x^n_\tau + a \textrm{sin}(1.5 \cdot (x^n_\tau-b)) + c
\end{align}
depends on the task parameters $\tau = (a,b,c,\alpha)$. The task parameters in \eqref{sinus} have the distributions $a \sim \mathcal{U}(0.9, 1.1), b \sim \mathcal{N}(0,0.06), c \sim \mathcal{N}(5,0.06),$ and $\alpha \sim \mathcal{N}(0.5,0.11)$.


\noindent \textbf{Schemes and Baselines.}
We compare the performance of the following schemes: 1) \textbf{Per-task learning} applies standard Bayesian learning on a per-task basis by minimizing the free energy \eqref{ELBO} over the variational parameters based only on training data for the given task. 2) \textbf{Joint learning} pools together the data samples from all tasks $\tau = 1,...,\mathcal{T}$ in order to optimize a single variational parameter  vector $\phi_\tau=\phi$ that minimizes the sum of the  free energy terms in \eqref{ELBO}  over all tasks $1,...,\mathcal{T}$. In a manner similar to the proposed meta-learning scheme, we allow for an additional step of fine-tuning of the variational parameters  using the training samples for the new, also known as meta-test, task $\tau^{*}$.  3) \textbf{Ideal per-task learning} provides a lower bound on the test error for the new task. It is obtained by optimizing the variational parameters to minimize \eqref{ELBO} by assuming access to an abundant data for the given task, with a tenfold increase as compared to the training data set size assumed for per-task learning, meta-learning and joint learning. 

\noindent \textbf{Architectures and Hyperparameters.} The Bayesian binary neural network comprises a single hidden layer with six neurons implementing rectified linear unit (ReLU) activation functions and a linear output neuron. The binary classifier used for density-ratio estimation (cf. \eqref{lossG}) has  two hidden layers with $256$ and $64$ hidden neurons, respectively, all implementing ReLU activation functions, with a sigmoid neuron for the output layer. As shown in Fig.~\ref{circ_one}, we use a hardware-efficient PQC comprised of 12 qubits, with a layer of Hadamard gates for state preparation, followed by $R_x$ and $R_z$ rotations, by an entangling layer of controlled-Z operations, and by a second layer of $R_x$ and $R_z$ rotations. 


Meta-learning is implemented over $240$ meta-training epochs, with $\mathcal{T} = 20$ meta-training tasks and task mini-batches of size $5$, unless stated otherwise. Optimization of the binary classifier uses $240$ training epochs, with mini-batches of size $10$, via the Adam optimizer with learning rate $0.01$. The number of samples for each task in the meta-training set is $N_\tau = 40$, with an equal number of training and testing samples. We also have $20$ training samples for the new, meta-task, task. 

\begin{figure}[tbp]
\centering
\includegraphics[width=0.99\linewidth]{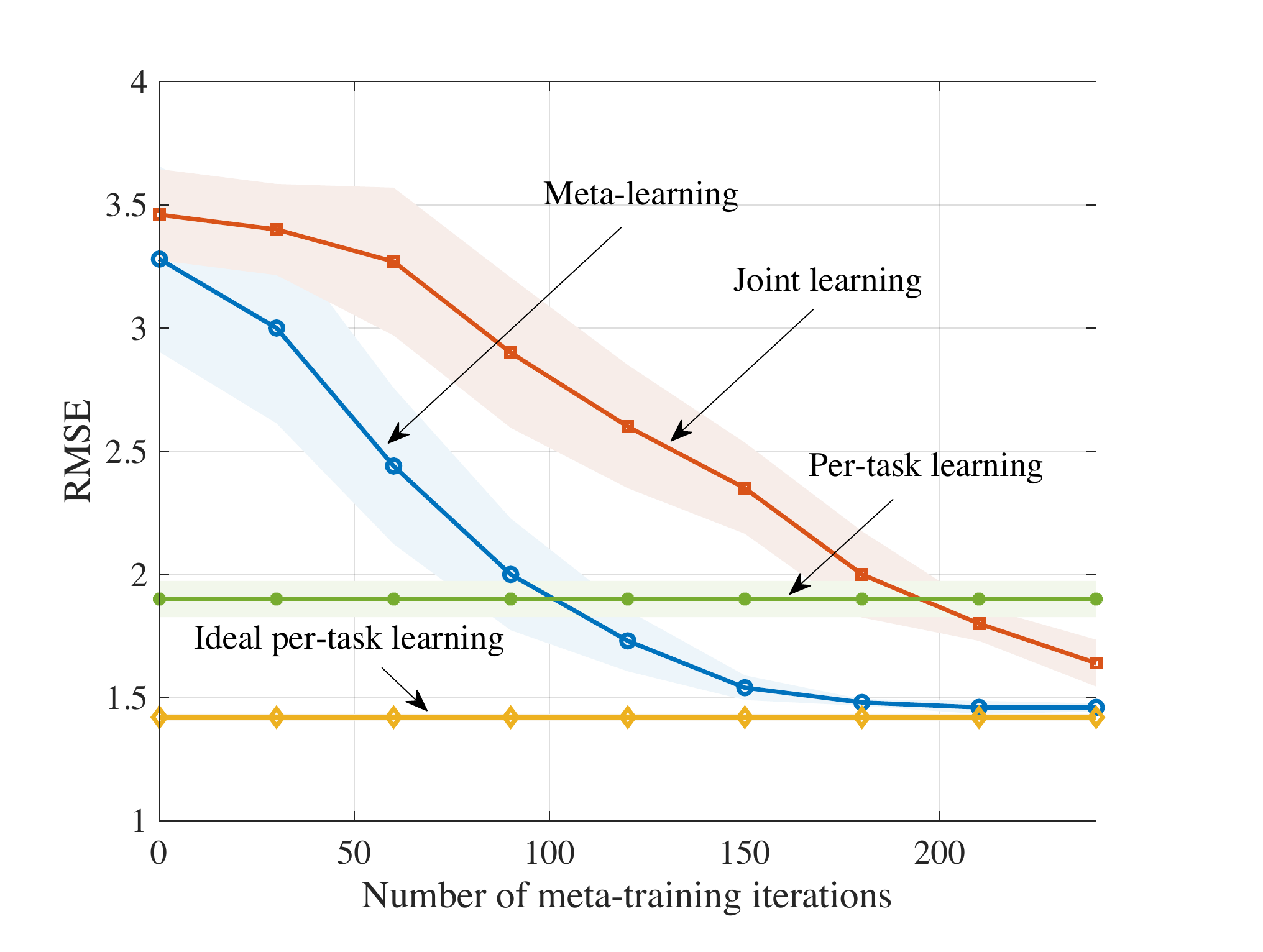}
\caption{Average RMSE for a new, meta-test, task as a function of the number of meta-training iterations.   The results are averaged over $5$ independent trials.
}
\vspace*{-5mm}
\label{rmse_one}
\end{figure}

\noindent \textbf{Results.} Fig.~\ref{rmse_one} illustrate the results in terms of the prediction root mean squared error (RMSE) as a function of the number of meta-training iterations. By comparison with conventional per-task learning, the figure illustrates the capacity of both joint learning and meta-learning to transfer knowledge from the meta-training to the meta-test task, with meta-learning clearly outperforming joint learning.  For example, meta-learning requires around $150$ meta-training iterations to achieve the same RMSE ideal per-task training, whilst joint-learning requires more than $200$ to achieve comparable performance. 

\begin{figure}[tbp]
\centering
\includegraphics[width=0.99\linewidth]{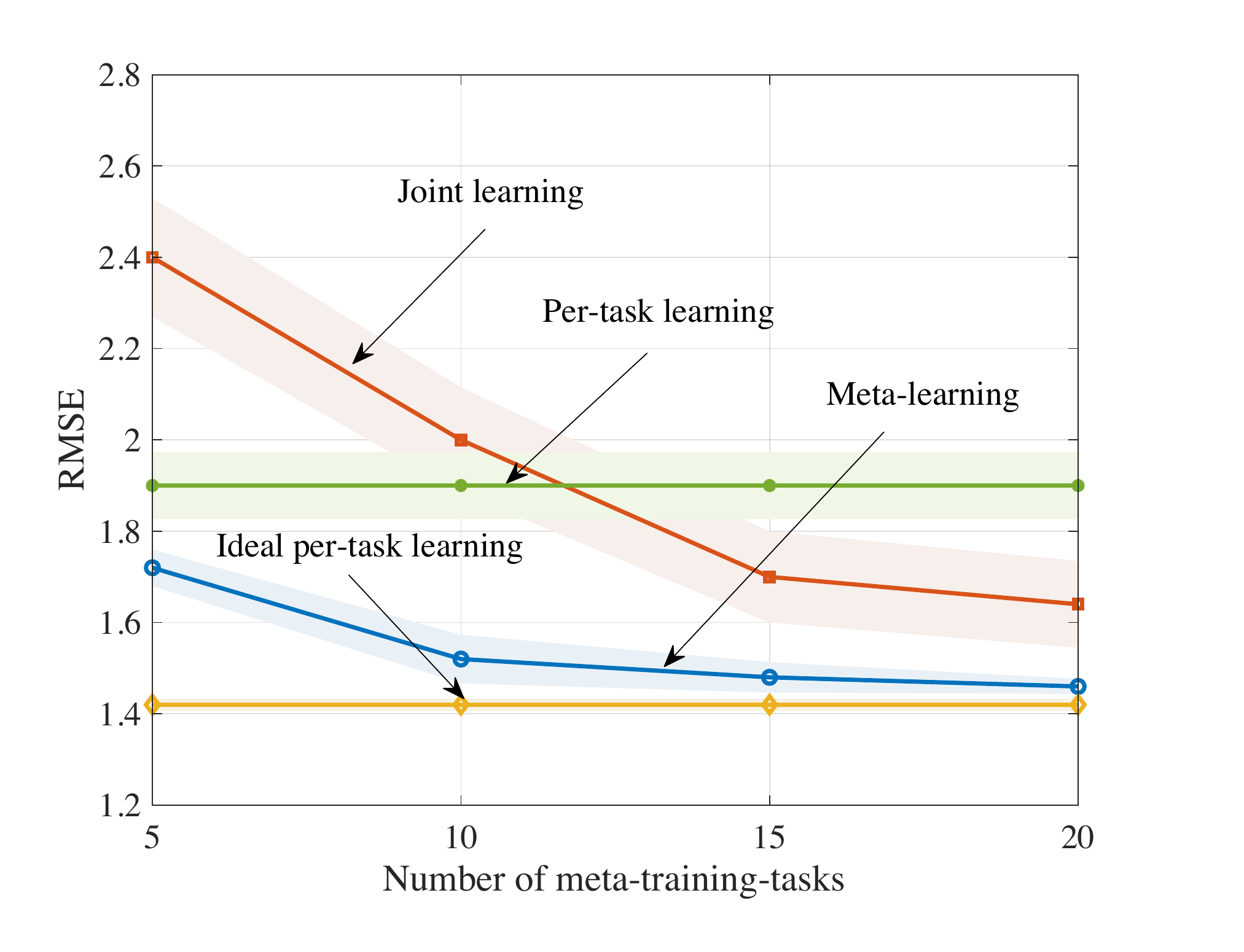}
\caption{Average RMSE for a new, meta-test, task as a function of the number of meta-training tasks. 
The results are averaged over $5$ independent trials.}
\vspace*{-5mm}
\label{rmse_two}
\end{figure}

Next, we investigate the effect of the number of meta-training tasks on the performance of joint learning and meta-learning by plotting the RMSE as a function of $\mathcal{T}$ in Fig.~\ref{rmse_two}.  The figure confirms that joint and meta-learning can outperform conventional per-task learning as long as one has access to data from a sufficiently large number of meta-training tasks. Furthermore, the gap in performance between joint and meta-learning is seen to become less prominent as the number of meta-training tasks increases, suggesting that meta-learning is most useful in low-data regimes. 

\section{Conclusions}
This paper has introduced a novel meta-learning method for Bayesian binary neural networks that leverages highly expressive variational posteriors implemented via parametric quantum circuits. Using gradient-based optimization of the parameters of the Born machine, the proposed method enables fast adaption to new learning tasks, outperforming conventional and joint learning strategies on a prototypical regression problem. Future work may consider more complex learning tasks and address a fully Bayesian formulation involving hyperpriors \cite{amit2018meta}.

\begin{appendices}

\section{Gradients for Parameter Updates}
The $j$th entry of the gradient for the inner update in \eqref{update_phi} can be computed as 
\begin{align}\label{gradG}
[\nabla_{\phi_\tau} \textrm{F}(\phi_\tau)]_j  
=& - \sum_{\theta_\tau} \frac{\partial}{\partial \phi_{\tau,j}} {q_{\phi_\tau}(\theta_{\tau,k})}   \log p(\mathcal{D}_\tau^\text{tr} \, | \, \theta_\tau)  \nonumber\\
& + \beta \,\, \sum_{\theta_\tau} \frac{\partial}{\partial \phi_{\tau,j}} {q_{\phi_\tau}(\theta_\tau)}  \text{logit} (d_{w}(\theta_\tau)),
\end{align}
where the partial derivatives are found using the parameter shift rule. Through this approach, the derivative of the expectation of any observable are calculated by evaluating the same circuit with a forward and backward shift of the argument \cite{schuld2018supervised}. This yields
\begin{align}\label{psr}
    \frac{\partial}{\partial \phi_{\tau,j}} {q_{\phi_\tau}(\theta_\tau)} = \frac{q_{\phi_\tau + s \textbf{e}_j}(\theta_\tau) - q_{\phi_\tau - s \textbf{e}_j}(\theta_\tau)}{2 \sin(s)},
\end{align}
where $s \neq k \pi, k \in \mathbb{Z}$, and $\textbf{e}_k$ is a one-hot vector with all zeros except for a 1 in position $j$. Thereby, for \eqref{gradG} we have 
\begin{align}\label{psh_first}
    [\nabla_{\phi_\tau} \textrm{F}(\phi_\tau)]_j =& \frac{1}{2 \sin(s)} \Big[-\E_{q_{\phi_\tau + s \textbf{e}_j}(\theta_\tau)}  [ \log p(\mathcal{D}_\tau^\text{tr} \, | \, \theta_\tau)\Big.]  \nonumber\\
    & \left. + \beta \,\, \E_{q_{\phi_\tau + s \textbf{e}_j}(\theta_\tau)} [ \text{logit} (d_{w}(\theta_\tau)) ] \right. \nonumber\\
    & \left. + \E_{q_{\phi_\tau - s \textbf{e}_j}(\theta_\tau)}   [\log p(\mathcal{D}_\tau^\text{tr} \, | \, \theta_\tau) \right.] \nonumber\\
    & \left. - \beta \,\, \E_{q_{\phi_\tau - s \textbf{e}_j}(\theta_\tau)} [ \text{logit} (d_{w}(\theta_\tau)) ] \right].
\end{align}

Updating the hyperparameter vector $\xi$ via \eqref{update_theta} is more involved, as it also requires differentiating through the update \eqref{update_phi} \cite{finn2017model}. To this end, one also needs to evaluate the Hessian matrix $\nabla^2_{\xi} \textrm{F}(\phi_\tau(\xi))$, which can be done via a double application of the parameter shift rule. This yields the $(j,l)$th element of the Hessian matrix as
\begin{align}\label{psh_second}
    [\nabla^2_{\xi} \textrm{F}(\phi_\tau(\xi))]_{j,l} =&\frac{1}{4 \sin(s_1) \sin(s_2)} 
     \Big[\E_{q_{\xi + s_1 \textbf{e}_j + s_2 \textbf{e}_l}(\theta_\tau)}   [f(\xi)] \Big. \nonumber\\
    & \left. - \E_{q_{\xi - s_1 \textbf{e}_j + s_2 \textbf{e}_l}(\theta_\tau)}   [f(\xi)] \right.  
     \nonumber\\
     &\left.-\E_{q_{\xi + s_1 \textbf{e}_j - s_2 \textbf{e}_l}(\theta_\tau)}   [f(\xi)]\right. 
    \nonumber\\
    &  \left.+ \E_{q_{\xi - s_1 \textbf{e}_j - s_2 \textbf{e}_l}(\theta_\tau)}   [f(\xi)] \right],
\end{align}
where $s_1, s_2 \neq k \pi, k \in \mathbb{Z}$, $\textbf{e}_j, \textbf{e}_l$ are one-hot vectors as defined above, and 
$f(\xi)=-\log p(\mathcal{D}_\tau^\text{tr} \, | \, \theta_\tau) +\beta \textrm{logit}(d_w(\phi_{\tau}))$. 
The expression in \eqref{psh_second} can be extended to an arbitrary number of inner loop updates by recursively applying the parameter shift rule.  For the presented experiments, we use $s_1 = s_2 = \pi/2$.


\end{appendices}



\bibliographystyle{IEEEbib}
\bibliography{litdab.bib}

\end{document}